\documentclass[nofootinbib,prd,twocolumn,showpacs,showkeys,preprintnumbers]{revtex4-1}
\usepackage{hyperref,amssymb,amsmath,mathrsfs,bm,graphicx}
\begin{document}
\title {Shear-free axially symmetric dissipative fluids}
\author{L. Herrera}
\email{lherrera@usal.es}
\affiliation{Escuela de F\'\i sica, Facultad de Ciencias, Universidad Central de Venezuela, Caracas, Venezuela}
\author{A. Di Prisco}
\email{adiprisc@ciens.ucv.ve}
\affiliation{Escuela de F\'\i sica, Facultad de Ciencias, Universidad Central de Venezuela, Caracas, Venezuela}
\author{J. Ospino}
\email{j.ospino@usal.es}
\affiliation{Departamento de Matem\'atica Aplicada and Instituto Universitario de F\'isica
Fundamental y Matematicas, Universidad de Salamanca, Salamanca, Spain}

\date{\today}
\begin{abstract}
We study the general properties of  axially symmetric dissipative configurations under the  shear-free condition. The link between the magnetic part of the Weyl tensor and the vorticity, as well as the role  of  the dissipative fluxes,  is clearly exhibited. As a particular case we examine the geodesic fluid. In  this latter case, the magnetic part of the Weyl tensor always vanishes, suggesting that no gravitational radiation is produced during the evolution. Also (for the geodesic case), in the absence of dissipation, the system evolves towards a FRW spacetime if the expansion scalar is positive. 
\end{abstract}
\pacs{04.40.-b, 04.40.Nr, 04.40.Dg}
\keywords{Relativistic Fluids, nonspherical sources, interior solutions.}
\maketitle

\section{Introduction}
In a recent paper \cite{1}, using a 1+3 approach \cite{21cil, n1, 22cil, nin}, we have developped a general framework for studying axially symmetric dissipative fluids.   In this work we endeavour to apply this approach to the specific case of shear-free fluids.

The relevance of the shear tensor in the evolution of self-gravitating systems and the consequences emerging from its vanishing have been discussed by many authors (see  \cite{SFCF}--\cite{s5} and references therein). 

 Furthermore as it has been recently shown \cite{exp} the  shear--free flow (in the nondissipative case)  appears to be equivalent to the well known homologous evolution. It should be recalled that   homology conditions are of great relevance in astrophysics  \cite{astr1}--\cite{astr3}.

Thus, in spite of the fact that the shear-free condition appears to be unstable with respect to some important physical phenomena \cite{ssf}, shear-free fluids play an important role in the study of self-gravitating objects.

As we shall see below the shear-free  condition brings out a clear link between the magnetic part of the  Weyl tensor  ($H_{\alpha \beta}$) and vorticity, even in the general, anisotropic and dissipative, case. It will be shown that for a shear-free fluid (not necessarily perfect), the necessary and sufficient condition to be irrotational is that the Weyl tensor be purely electric, thus generalizing a result  by Barnes \cite{b1,b2} and Glass \cite{glass}.

The subcase represented by the geodesic fluid is analyzed in some detail, in particular the dissipationless case. In this latter case it is shown that if the expansion scalar is positive, the system relaxes asymptotically to a FRW spacetime. Also, it is shown that the magnetic part of the Weyl tensor always vanishes in this case

In order to avoid rewriting most of the equations, we shall very often  refer  to \cite{1}.  Thus, we  suggest  that  the reader have at hand reference \cite{1}, when reading this manuscript.

\section{The shear-free condition and its consequences}
We shall consider  axially and reflection symmetric fluid distributions (not necessarily bounded). For such a system the most general line element may be written in ``Weyl spherical coordinates'' as:

\begin{equation}
ds^2=-A^2 dt^2 + B^2 \left(dr^2
+r^2d\theta^2\right)+C^2d\phi^2+2Gd\theta dt, \label{1b}
\end{equation}
where $A, B, C, G$ are positive functions of $t$, $r$ and $\theta$. We number the coordinates $x^0=t, x^1=r, x^2= \theta, x^3=\phi$.

The energy momentum tensor in the ``canonical'' form reads:
\begin{eqnarray}
{T}_{\alpha\beta}&=& (\mu+P) V_\alpha V_\beta+P g _{\alpha \beta} +\Pi_{\alpha \beta}+q_\alpha V_\beta+q_\beta V_\alpha,
\label{6bis}
\end{eqnarray}
where as usual, $\mu, P,  \Pi_{\alpha \beta}, V_\beta, q_\alpha$ denote the eneregy density, the isotropic pressure, the anisotropic stress tensor, the four velocity and the heat flow vector, respectively.
The anisotropic stress tensor  may be written in terms of  three scalar functions ($\Pi_I, \Pi_{II}, \Pi_{KL})$, whereas the heat flow vector is defined by two scalar functions  $q_I, q_{II}$ (see eqs. (10-16) in \cite{1} for details).

The shear tensor is defined by two scalar functions $\sigma_I, \sigma_{II}$, which in terms of the metric functions read (see eqs.(20-25) in \cite{1}):

\begin{eqnarray}
2\sigma_I+\sigma_{II}&=&\frac{3}{A}\left(\frac{\dot B}{B}-\frac{\dot C}{C}\right) \label{primers}\\
2\sigma_{II}+\sigma_I&=&\frac{3}{A^2B^2r^2+G^2}\,\left[AB^2r^2\left(\frac{\dot B}{B}-\frac{\dot C}{C}\right)\right.\nonumber\\
&
&\left.+\frac{G^2}{A}\left(-\frac{\dot A}{A}+\frac{\dot G}{G}-\frac{\dot C}{C}\right)\right]\label{sigmas}
\end{eqnarray}

For the other kinematical variables (the expansion, the four acceleration and the vorticity) we have:

The expansion

\begin{eqnarray}
\Theta&=&\frac{AB^2}{r^2A^2B^2+G^2}\,\left[r^2\left(2\frac{\dot B}{B}+\frac{\dot C}{C}\right)\right.\nonumber\\
&&+\left.\frac{G^2}{A^2B^2}\left(\frac{\dot B}{B}-\frac{\dot A}{A}+\frac{\dot G}{G}+\frac{\dot C}{C}\right)\right].
\label{theta}
\end{eqnarray}

The four acceleration
\begin{equation}
a_\alpha=V^\beta V_{\alpha;\beta}=a_I K_\alpha+a_{II}L\alpha,\label{acc}
\end{equation}
\noindent with vectors $\bold K$ and $\bold L$  having components:
\begin{equation}
K_\alpha=(0, B, 0, 0);\qquad L_\alpha=(0, 0, \frac{\sqrt{A^2B^2r^2+G^2}}{A}, 0),\label{vec}
\end{equation}

\noindent and where the  two scalar functions ($a_I, a_{II}$) are defined by (see eq.(17) in \cite{1})
 \begin{eqnarray}
 a_I&=& \frac {A^{\prime} }{AB},
 \\
 a_{II}&=&\frac{A}{\sqrt{A^2B^2r^2+G^2}}\left[\frac{G}{A^2}\left(-\frac {\dot A}{A}+\frac {\dot G}{G}\right)+\frac {A_{,\theta}} {A}\right],
\label{acc'}
\end{eqnarray}
whereas the vorticity vector is defined through a single scalar $\Omega$, given by (see eq.(29) in \cite{1})

\begin{equation}
\Omega =\frac{(AG^\prime-2GA^\prime)}{2AB\sqrt{A^2B^2r^2+G^2}}\label{omega},
\end{equation}
where  primes and dots denote derivatives with respect to $r$ and $t$ respectively.

\noindent If we assume the evolution to be shear-free, i.e.
\begin{equation}
\sigma _I=\sigma _{II}=0,\label{sfc}
\end{equation}
\noindent then from (\ref{primers}) and (\ref{sigmas}) we have
\begin{eqnarray}
C(t,r,\theta)=R(r,\theta)B(t,r,\theta)\nonumber
\\
G(t,r,\theta)=A(t,r,\theta)B(t,r,\theta)\tilde G(r,\theta).\label{sfcon}
\end{eqnarray}
From regularity conditions at the origin we must require $R(0,\theta)=\tilde G(0,\theta)=0$.

Next,  from  (A.5) in \cite{1} we may write, if $\sigma_{\alpha\beta}=0$,
\begin{equation}
\nabla_{\langle \alpha} \omega_{\beta\rangle}+2\omega_{\langle\alpha }a_{ \beta \rangle}=H_{\alpha \beta},
\label{ne1}
\end{equation}
where angled brackets denote the spatially projected, symmetric and trace-free part, and $\nabla_\alpha  \omega_\beta \equiv h^\delta_\alpha \omega_{\beta;\delta}$.

From the above it follows at once  that $\omega_\alpha=0 \Rightarrow H_{\alpha \beta}=0$. Furthermore the inverse is also true.  Indeed, assuming $H_{\alpha \beta}=0$ in (\ref{ne1}), we obtain 
\begin{equation}
\nabla_\alpha \omega^\alpha=-2a_\alpha \omega^\alpha,
\label{ne2}
\end{equation}
however, if the shear tensor vanishes, the following identity holds
\begin{equation}
\nabla_\alpha \omega^\alpha=a_\alpha \omega^\alpha.
\label{ne3}
\end{equation}
Eqs. (\ref{ne2}) and (\ref{ne3}) imply that $\omega_{\alpha \beta}=0$. Alternatively,   the regularity condition at the origin $\omega_\alpha(r=0)=0$, can be analytically extended to the whole distribution, by taking successive $\nabla_\alpha$ derivatives of (\ref{ne2}), thereby leading  to the same result.

Thus, using the notation of \cite{1}, we have established that
\begin{equation}
 H_1=H_2=0\Leftrightarrow \Omega =0, \label{ne4}
\end{equation}
where $H_1$ and $H_2$ are the two scalar functions which define the magnetic part of the Weyl tensor.

It is important to stress the point that in order to arrive at (\ref{ne4}) we have used the tensorial equation (A.5) (eq.(\ref{ne1}) above), which is not restricted to the axially symmetric case.
In other words the necessary and sufficient condition for a shear-free fluid to be  irrotational is that the Weyl tensor be purely electric. This generalizes a result by Barnes \cite{b1, b2} and Glass \cite{glass}, to anisotropic and dissipative fluids (observe that in \cite{1} it was incorrectly stated that such a generalization only applies to non dissipative fluids).

\noindent For the heat flow scalars we obtain in this case (shear--free and  axially symmetric), using (B.6) and (B.7) from \cite{1}
\begin{eqnarray}
4\pi q_I &=& \frac{1}{3B}\Theta^{\prime},\label{inho1}
\\
4\pi q_{II} &=& \frac{1}{3B r}\Theta _{,\theta}.\label{inho2}
\end{eqnarray}

Thus in the dissipationless case, the expansion scalar is homogeneous, $\Theta=\Theta(t)$.
\section{Geodesic Condition: $a_\alpha=0$}
We shall further restrict our system to the case of vanishing four--acceleration. Two important observations are in order at this point:
\begin{itemize}
\item As it will be shown below, all, geodesic and  shear--free fluids, are necessarily irrotational.
\item Shear--free irrotational, geodesic fluids, have been analyzed in great detail by Coley and McManus \cite{c1, c2}. Here we look at the axially symmetric heat conducting case of these fluid
distributions.
\end{itemize}
\noindent Next, the geodesic condition  implies that

\begin{equation}
a_I=\frac{A^\prime}{AB}=0\,\,\,\Rightarrow \,\,\, A=\tilde A(t,\theta)\label{a1}
\end{equation}
\noindent and
\begin{equation}
a_{II}=\frac{1}{B\sqrt{r^2+\tilde G^2}}(\frac{\tilde G\Theta B}{3}+\frac{A_{,\theta}}{A})=0\,\,\,\Rightarrow\,\,\, \tilde G \Theta B=F_2(t,\theta)\label{a2}
\end{equation}
\noindent Given that $\Omega (t,0,\theta)=\tilde G(t,0,\theta)=0$, from (\ref{a2}) we find that
\begin{equation}
F_2(t,\theta)=0 \,\,\,\Rightarrow\,\,\,\Omega=0 \,\,\, {\rm or} \,\,\, \Theta=0.
\end{equation}
The above results can also be obtained from (A.3) in \cite{1}, which reads in this particular case as:
\begin{equation}
h^\beta_\alpha V^\delta \omega_{\beta;\delta}=-\frac{2}{3}\Theta \omega_\alpha.
\label{ne5}
\end{equation}

Indeed, combining the above equation or its projection on the ${\bold {KL}}$ vectors (eq. (B.5) in \cite{1}), with (\ref{primers}), (\ref{sigmas}) and (\ref{theta}) we obtain the same result, i.e. $\Theta \Omega=0$. This is in agreement with the so called ``shear--free conjecture'' for perfect fluids, which suggests that $\sigma_{\alpha \beta}=0$ implies $\Theta \Omega=0$ (see \cite{vb} and references therein). Here we have not restricted ourselves to the perfect fluid case, although our result only applies to geodesic fluids.

Let us first consider the case:
\subsection{\quad $\Omega _{\alpha \beta}=0, \Theta \neq 0.$}
In this case the line element takes the form
\begin{equation}
ds^2=-dt^2+B^2(t,r,\theta)\left[dr^2+r^2d\theta ^2+R^2 (r,\theta)d\phi ^2\right],
\end{equation}
and the following equations have to be satisfied:

The ``continuity''  equation (Eq.(A.6) in \cite{1})
\begin{equation}
\dot \mu +(\mu+P)\Theta+q^\alpha _{;\alpha}=0.\label{muq}
\end{equation}
The generalized ``Euler'' equation (Eq.(A.7) in \cite{1}) 
\begin{equation}
h^\beta _{\alpha}(P_{,\beta}+\Pi^\mu_{\beta;\mu}+q_{\beta;\mu}V^\mu)+\frac{4}{3}\Theta q_\alpha=0.\label{ecq}
\end{equation}
In the non-dissipative case, it is known that the shear-free condition poses restrictions on the equilibrium equation of state (see \cite{SFCF, s1}) even in the non-geodesic case. Thus, it is legitimate  to ask whether or not, in our case, any admissible equation of state is restricted by the transport equation assumed for the heat transport? 
 
The answer to the above question seems to be affirmative, if we  observe that the last term within the round bracket in (\ref{ecq}), is related to the thermodynamic variables through the transport equation (57) in \cite{1} (if we assume the Israel-Stewart   theory). However, in the  general case ($a_\alpha \ne 0$), there is  four--acceleration--heat coupling, and the answer is not so evident, requiring a more detailed analysis which is  outside of the scope of this manuscript. 

Next, from equations (B.2, B.3, B.4) in \cite{1}, it follows that
\begin{equation}
Y_I=Y_{KL}=Y_{II}=0,\label{losY}
\end{equation}
implying that  
\begin{eqnarray}
X_I=-2{\cal E}_I, X_{II}=-2{\cal E}_{II}, X_{KL}=-2{\cal E}_{KL},\\ \nonumber 
{\cal E}_I=4\pi \Pi_I, {\cal E}_{II}=4\pi \Pi_{II}, {\cal E}_{KL}=4\pi \Pi_{KL}, 
\label{losx}
\end{eqnarray}
where ${\cal E}_{I}, {\cal E}_{II}, {\cal E}_{KL}$ are the three scalar functions defining the electric part of the Weyl tensor (see \cite{1} for details), and $Y_I, Y_{KL}, Y_{II}, X_I, X_{II}, X_{KL}$ are some of the structure scalars obtained from the orthogonal splitting of the Riemann tensor which are  defined in eqs.(38-50) in \cite{1}.

\noindent Also, as  stated before, from (B.6) and (B.7), we obtain (\ref{inho1}) and (\ref{inho2}).

Finally, B.10--B.18 in \cite{1}, produce  (some of which are redundant):
\begin{equation}
-\frac{1}{3}(X_I-4\pi\mu \dot ) +\frac{1}{3}\mathcal {E}_I\Theta=-\frac{4\pi}{3}(\mu+P+\frac{1}{3}\Pi _I)\Theta-\frac{4\pi}{B}q^\prime_{I}-4\pi  \frac{q_{II}B_{\theta}}{B^2 r},\label{A10}
\end{equation}

\begin{equation}
-\dot X_{KL}-\Theta X_{KL}=\frac{8\pi}{3}\Pi_{KL}\Theta-2\pi(K^\mu L^\nu+K^\nu L^\mu)q_{\nu;\mu},\label{NUEVA}
\end{equation} 

\begin{equation}
\frac{1}{3}(-X_{II}+4\pi\mu \dot )+\frac{\Theta}{3}\mathcal {E}_{II}=-\frac{4\pi}{3}(\mu+P+\frac{1}{3}\Pi _{II})\Theta -4\pi L^\mu L^\nu q_{\nu;\mu}.\label{A12}
\end{equation}

We shall now specialize to the dissipationless case  $q_I=q_{II}=0$, (similar  to the models analyzed in \cite{c2} although the anisotropic stress tensor is more general). The Petrov type of each specific model depending on the number of distinct eigenvalues of $\Pi_{\alpha \beta}$ \cite{t, kramer}.

\noindent From the equations (\ref{theta}), (\ref{inho1}), (\ref{inho2}), (\ref{muq}) and (\ref{losY}) it follows at once that in the dissipationless case:
\begin{eqnarray}
\Theta &=&\Theta(t) \,\,\Rightarrow \,\,B(t,r,\theta)=f(t)b(r,\theta),\nonumber
\\
 \mu &= &\mu(t) \,, P=P(t)\, , \Pi_I=\Pi_I(t),\,\,\\ \nonumber
\Pi_{II}&=&\Pi_{II}(t) \,, \Pi_{KL}=\Pi_{KL}(t),
\end{eqnarray}

\noindent  where the fact has been used that $Y_T=4\pi(\mu+3P)$ (Eq.(42) in \cite{1}).

Then, equations (\ref{A10}), (\ref{NUEVA}) and (\ref{A12}) may be easily integrated to obtain
\begin{eqnarray}
{\cal E}_I&=&{\cal E}_I(0)exp[-\frac{2}{3}\int{\Theta dt}] ,\qquad {\cal E}_{II}={\cal E}_{II}(0)exp[-\frac{2}{3}\int{\Theta dt}],\nonumber  \\ {\cal E}_{KL}&=&{\cal E}_{KL}(0)exp[-\frac{2}{3}\int{\Theta dt}],\label{integra}
\label{int}
\end{eqnarray}
or feeding the expression of $\Theta$ back into (\ref{integra}):
\begin{equation}
{\cal E}_I=\frac{{\cal E}_I(0)}{B^2},\qquad {\cal E}_{II}=\frac{{\cal E}_{II}(0)}{B^2},\qquad {\cal E}_{KL}=\frac{{\cal E}_{KL}(0)}{B^2}. \label{nint}
\end{equation}
From the above it becomes evident that $B=f(t)$, and in the case $\Theta>0$ the system tends to a FRW spacetime.

Let us now analyze the other case.
\subsection{\quad $\Theta =0, \Omega \neq 0.$}
In this case the system becomes time independent, as it can be easily inferred from (\ref{theta}).

Then, from B.1, B.2, B.3 and B.4 in \cite{1}, we obtain:
\begin{equation}
2\Omega^2=Y_T=2 Y_I=2 Y_{II}, \qquad Y_{KL}=0,
\label{notra}
\end{equation}

and  from B.6 and B.7  in \cite{1} :
\begin{equation}
-(\Omega BR)_{,\theta}=8\pi q_IB^2R\sqrt{r^2+\tilde G^2},\label{nq1n}
\end{equation}

\begin{equation}
(\Omega BR)^\prime=8\pi q_{II}B^2 R. \label{nq2n}
\end{equation}

From the two equations above it becomes evident that in the dissipationless case, $\Omega=0$, implying because of (\ref{notra}), that $\mu=P=0$, unless we assume the equation of state $\mu=-3P$. In other words, any model belonging to this class ($\Theta=0$) must  necessarily  be dissipative.

However it is a simple matter to check, from  B.8,  B.9  and B.13 in \cite{1}, together with   (\ref{nq1n}), (\ref{nq2n}) and the regularity conditions on the axis of symmetry, that   no such models ($\Theta=0$) exist.
 
\section{conclusions}
Using the framework developed in \cite{1} we have analyzed in some detail the general properties of the shear-free case. We have seen that, for a general dissipative and anisotropic fluid, vanishing vorticity, is a  necessary and sufficient condition for the magnetic part of the Weyl tensor  to vanish, providing a generalization of  the same result for perfect fluids obtained in \cite{b1, b2, glass}.
This result, in turn, implies  that  vorticity should necessarily appear if the system  radiates gravitationally. We stress that this result is not restricted to the axially (and reflection) symmetric case. This further reinforces  the well established  link between radiation and vorticity ( see \cite{link} and references therein).

In the geodesic case,  the vorticity always  vanishes (and thereof  the magnetic parts of the Weyl tensor), suggesting that in this latter case no gravitational radiation is produced  during the evolution.  This result is in agreement with the shear--free conjecture mentioned above.  However we do not know if it holds for the non--geodesic case. if it does, then it is clear that we should consider shearing fluids, when looking for sources of gravitational radiation.

The above  result is also similar to the one obtained for the cylindrically symmetric case \cite{cyl}, and suggests (as does the shear--free conjecture) a link between the shear of the source and the generation of gravitational radiation during the evolution.

In the geodesic case we also observe that,  in the non-dissipative case, the models do not need to be FRW (as already stressed in \cite{c1}), however  the system relaxes to the FRW spacetime  (if $\Theta$ is positive). In presence of  dissipative fluxes, such tendency does not appear,  further illustrating the relevance of dissipative processes in the evolution of self-gravitating fluids.


\begin{thebibliography}{100}
\bibitem{1} L. Herrera, A. Di Prisco, J. Ib\'a\~nez and  J. Ospino,
{\it Phys.  Rev. D} {\bf  89}, 084034, (2014).
\bibitem{21cil} G. F. R. Ellis {\ Relativistic Cosmology} in: Proceedings of the International School of Physics `` Enrico Fermi'', Course 47: General Relativity and Cosmology. Ed. R. K. Sachs (Academic Press, New York and London) (1971).
\bibitem{n1} G. F. R. Ellis and H.  van Ellst, {\it gr--qc/9812046v4} (1998).
\bibitem{22cil} G. F. R. Ellis {\it Gen. Rel. Grav.} {\bf  41}, 581 (2009).
\bibitem{nin} G. F. R. Ellis, R. Maartens and M. A. H. MacCallum, {\it Relativistic Cosmology} (Cambridge U. P., Cambridge) (2012).
\bibitem{SFCF} C.  B. Collins and J. Wainwright,  {\it Phys. Rev. D}  {\bf 27},
 1209 (1983).
\bibitem{s1}  E.  N.  Glass,  {\it J. Math. Phys.}  {\bf 20}, 1508 (1979).
\bibitem{s2} R. Chan, {\it Mon. Not. R. Astron. Soc.} {\bf 299}, 811 (1998).
\bibitem{s3p} P. Joshi, N. Dadhich and R. Maartens, {\it Phys. Rev. D} {\bf 65}, 101501 (2002).
\bibitem{s3}  P. Joshi, R.  Goswami and N. Dadhich, {\it gr--qc}/0308012.
\bibitem{s4} L.  Herrera and N. O. Santos, {\it Mon. Not. R. Astron. Soc.}  {\bf 343},
 1207 (2003).
\bibitem{s5} M. Govender, K. P. Reddy and S. D. Maharaj, {\it Int. J. Mod. Phys. D} {\bf 23}, 1450013 (2014).
\bibitem{exp} L. Herrera, N. O. Santos and A. Wang, {\it Phys. Rev. D} {\bf 78}, 084026 (2008).

\bibitem{astr1}  M. Schwarzschild, {\it Structure and Evolution
of the Stars}, (Dover, New York) (1958).
\bibitem{astr2} R. Kippenhahn and A. Weigert,
{\it Stellar Structure and Evolution}, (Springer Verlag, Berlin) (1990).
\bibitem{astr3} C.
Hansen and S. Kawaler, {\it Stellar Interiors: Physical Principles,
Structure and Evolution}, (Springer
Verlag, Berlin) (1994).
\bibitem{ssf} L. Herrera, A. Di Prisco and  J. Ospino, {\it Gen. Relativ. Gravit.}  {\bf 42}, 1585 (2010).
\bibitem{b1} A.  Barnes, {\it Gen Rel. Gravit} {\bf 4}, 105 (1973).
\bibitem{b2}A.  Barnes, in {\it Classical general
relativity}. Eds. W.B Bonnor, J.N. Islam and M.A.H. MacCallum (Cambridge
University Press) (1984).
\bibitem{glass} E. N. Glass,  {\it J. Math. Phys.} {\bf 16}, 2361 (1975).
\bibitem{c1}  A. A. Coley and D. J. McManus, {\it Class. Quantum Grav.} {\bf 11}, 1261 (1994).
\bibitem{c2}   D. J. McManus and A. A. Coley, {\it Class. Quantum Grav.} {\bf 11}, 2045 (1994).
\bibitem{vb} N Van den Bergh, J. Carminati and H. R. Karimian, {\it Class. Quantum Grav.} {\bf 24}, 3735 (2007).
\bibitem{t} M. Trumper, {\it J. Math. Phys.} {\bf 6}, 584 (1965).
\bibitem{kramer} H. Stephani, D. Kramer, M. MacCallum, C. Honselaers and
E. Herlt, {\it Exact Solutions to Einstein's Field Equations. Second
Edition}, (Cambridge University Press, Cambridge), (2003).
\bibitem{link} L. Herrera, {\it Gen. Relativ. Gravit.} {\bf 46}, 1654 (2014).
\bibitem{cyl}A. Di Prisco, L. Herrera, M. A. H. MacCallum and N. O. Santos, {\it Phys. Rev. D} {\bf 80}, 064031 (2009).
\end{thebibliography}
\end{document}